\documentclass[12pt]{article}

\RequirePackage[OT1]{fontenc}
\usepackage{amsmath}
\usepackage{mathabx}
\usepackage{graphicx,psfrag,epsf}
\usepackage{enumerate}
\usepackage{natbib}
\usepackage{url} % not crucial - just used below for the URL
\usepackage{appendix}

\newcommand{\blind}{1}

\usepackage{amsthm}
\usepackage[hidelinks]{hyperref}
\usepackage{setspace}
\usepackage{pdflscape}
\usepackage{afterpage}
\usepackage[top=1in,right=1.25in,left=1.25in,bottom=1in]{geometry}

\usepackage{url} % not crucial - just used below for the URL
\usepackage{threeparttable}
\usepackage{booktabs}
\usepackage{multirow}
\usepackage{bm}

\newcommand{\specialcell}[2][c]{%
  \begin{tabular}[#1]{@{}c@{}}#2\end{tabular}}

\theoremstyle{plain}

\newtheorem{assumption}{Assumption}[section]
\begin{document}

\def\spacingset#1{\renewcommand{\baselinestretch}%
{#1}\small\normalsize} \spacingset{1}

%%%%%%%%%%%%%%%%%%%%%%%%%%%%%%%%%%%%%%%%%%%%%%%%%%%%%%%%%%%%%%%%%%%%%%%%%%%%%%

\if1\blind
{
  \title{\bf Blending of Probability and Non-Probability Samples: Applications to a Survey of Military Caregivers}
  \author{\large Michael W. {Robbins}, %\thanks{Corresponding author: Tel: 1-412-683-2300 ext.~4914, Email: mrobbins@rand.org},
Bonnie {Ghosh-Dastidar} and Rajeev {Ramchand}\thanks{Michael W.~Robbins is Statistician with the RAND Corporation, Pittsburgh, PA 15213 (E-mail:~{\em mrobbins@rand.org}).  Bonnie Ghosh-Dastidar is Senior Statistician (E-mail:~{\em bonnieg@rand.org}) and Rajeev Ramchand is Senior Behavioral and Social Scientist (E-mail:~{\em ramchand@rand.org}) with the RAND Corporation, Arlington, VA 22202.  This work was partially funded by the Elizabeth Dole Foundation and by National Science Foundation (DIIS/BIGDATA) Award \#1837959.  The views expressed are those of the authors and do not necessarily represent the views of RAND, the Elizabeth Dole Foundation, or the NSF.}
}
  \maketitle
} \fi

\if0\blind
{
  \bigskip
  \bigskip
  \bigskip
  \begin{center}
    {\LARGE\bf Blending of Probability and Non-Probability Samples: Applications to a Survey of Military Caregivers}
\end{center}
  \medskip
} \fi

\bigskip
\begin{abstract}
Probability samples are the preferred method for providing inferences that are generalizable to a larger population. However, when a small (or rare) subpopulation is the group of interest, this approach is unlikely to yield a sample size large enough to produce precise inferences. Non-probability (or convenience) sampling often provides the necessary sample size to yield efficient estimates, but selection bias may compromise the generalizability of results to the broader population. Motivating the exposition is a survey of military caregivers; our interest is focused on unpaid caregivers of wounded, ill, or injured servicemembers and veterans who served in the US armed forces following September 11, 2001.  An extensive probability sampling effort yielded only 72 caregivers from this subpopulation. Therefore, we consider supplementing the probability sample with a convenience sample from the same subpopulation, and we develop novel methods of statistical weighting that may be used to combine (or blend) the samples. Our analyses show that the subpopulation of interest endures greater hardships than caregivers of veterans with earlier dates of service, and these conclusions are discernably stronger when blended samples with the proposed weighting schemes are used. We conclude with simulation studies that illustrate the efficacy of the proposed techniques, examine the bias-variance trade-off encountered when using inadequately blended data, and show that the gain in precision provided by the convenience sample is lower in circumstances where the outcome is strongly related to the auxiliary variables used for blending.
\end{abstract}

\noindent%

\newtheorem{proposition}{Proposition}
\newtheorem{lemma}{Lemma}
\newtheorem{theorem}{Theorem}
\newtheorem{corollary}{Corollary}
\def\convD{\,{\buildrel D \over\to}\,}
\def\convP{\,{\buildrel p \over\to}\,}
\def\equalD{\,{\buildrel D \over =}\,}

%\linenumbers
%\doublespacing
\def\spacingset#1{\renewcommand{\baselinestretch}%
{#1}\small\normalsize} \spacingset{1.5}

\section{Introduction}\label{sec1}

Probability (or representative) sampling, in which each member of a population is sampled with a known and nonzero probability, has long served as the gold-standard for producing inferences that are generalizable to the whole population.  However, researchers are increasingly interested in deriving estimates on small segments of a population, and it is often infeasible to use probability sampling to produce samples of sufficient size from such segments due to the lack of prevalence of individuals therein.

One small segment of the population for which there is great interest is service members and veterans who served in US military operations in Iraq and Afghanistan since 2001. Policymakers are increasingly interested in identifying the health, educational, labor, and other social needs of this population so that they can form policies or create programs to better address these needs.  In 2014, there were 2.6 million post-9/11 veterans, representing 0.8\% of the total US population.  Although the number of veterans is expected to increase to 3.6 million in 2019, it is still projected to only comprise 1.0\% of the population \citep{NCVAS15}.
Oftentimes, researchers are interested in even further narrowly defined segments of veterans. Specifically, this article is motivated by efforts to draw inferences regarding the wellbeing of unpaid caregivers of wounded, ill, and injured service members veterans of the US armed forces who served after September 11, 2001 (post-9/11) and caregivers of those who did not (pre-9/11).
Sampling of veterans of the Armed Forces is aided by databases that are compiled by organizations that provide services to such individuals (e.g., Iraq and Afghanistan Veterans of America, Disabled Veterans of America, American Legion).  Utilization of these services is voluntary; therefore, these databases do not yield sampling frames that are representative of the veteran population, and therefore samples drawn using these lists are considered non-probability samples.

Non-probability (or convenience) samples are becoming an increasingly efficient and cost-effective alternative to probability samples.
\cite{baker13a} provide an outline of types of convenience samples (e.g., mall intercepts, volunteer panels, river samples and respondent driven samples).  Of particular interest in the current literature are volunteer (opt-in) web panels \citep[e.g.,][]{baker10, biffignandi12, tourangeau13}.
Convenience samples are usually not representative of the complete population from which they were drawn, and inferences extracted from them do not necessarily generalize to that population (i.e., such samples may be fraught with selection bias).
Therefore, a wealth of literature \citep[e.g.,][]{schonlau04, duffy05, schonlau07,  chang09, blasius10, yeager11} has been devoted to evaluating the quality of inferences drawn from convenience samples.

It is possible to use non-representative samples to produce estimates that generalize to a population even in circumstances where micro-level data that are representative of that population are not available.  For instance, if the non-representative sample is differentiated from the population solely across characteristics for which aggregate measures are available for the entire population, a host of statistical weighting techniques, wherein certain observations are provided greater emphasis than others in estimator calculations, may be applicable \citep[see][for example]{wang15}.  However, it is often the case that no comprehensive summary quantities are available from the entire segment of interest or that the population and convenience sample are differentiated on the basis of subtle characteristics that are not commonly reported in census-type summaries.  These disadvantages can be overcome by using a (potentially small) probability sample in tandem with non-representative data.

The idea of combining data from two or more surveys has a lengthy history within extant literature \citep[e.g.,][]{zieschang86, renssen97, merkouris04, rao10}.  This discussion has spurned the development of methods for combining (or blending) data from probability and convenience samples so as to yield generalizable inferences.  \cite{elliott07} and \cite{ghosh09} discuss composite estimators---i.e., estimation involving outcomes is performed on each sample individually and the results are combined in some manner.  Here, we focus on weighting techniques that enable the two datasets to be analyzed as a single dataset (following weighting).
We separate these methods into two general classes of procedures:~1) propensity score methods \citep{lee06, lee09, schonlau09, valliant11}, which model the mechanism that differentiates the two samples, and 2) calibration weighting \citep{lee09, disogra11}, which determines weights that satisfy the constraint that totals calculated across a set of variables using a weighted sample equal corresponding totals calculated using the entire population.
Each of these weighting procedures is designed to match the convenience sample to the population across a set of auxiliary variables; congeniality between the sample and the remainder of the population across other variables (e.g., outcomes) is assumed to hold following weighting adjustments.

Our first methodological contribution is to introduce a novel and easily applicable manner of using propensity scores to calculate weights by directly approximating inclusion probabilities.  We rigorize the framework for combining samples by developing and distinguishing between methods that weight the samples separately (wherein each sample is weighted to be individually representative of the population) and simultaneously (in which the samples are only representative when used in tandem), and we illustrate the comparative advantages of these two types of weights.
Specifically, weights based on disjoint blending enable evaluation of the representativeness of the blended sample through statistical testing of the adequacy of blending (i.e., are there latent characteristics that differentiate the two samples?).
In contrast, weights based on simultaneous blending appear to yield smaller variances than their disjoint counterparts and do not require that the convenience sample covers the population.

For the application discussed herein, we utilize data from a survey administered as part of an extensive study of military caregivers; the survey and a wealth of findings from the study are described in detail in \cite{ramchand14}.
The caregivers survey was administered to a probability-based panel of approximately 45,000 American adults (GfK KnowledgePanel); however, this yielded only 72 post-9/11 caregivers.
A convenience sample that provided an additional 281 post-9/11 military caregivers was derived from a database provided by the Wounded Warrior Project (WWP).
Response patterns from caregivers sampled from the WWP are markedly different from the those given by respondents in the probability sample.  Nonetheless, the proposed methods for weighting are shown to be effective at combining data from the two sources.
Using only the probability sample, we estimate that post-9/11 caregivers score 1.5 points higher average than pre-9/11 caregivers on a 24-point measure of depression levels.  However, a 95\% confidence interval on this estimate ranges from -0.36 to 3.38 points.  Using the blending methods outlined here, we see a marked improvement in precision of this estimate (see Section \ref{sec2} for details).

The weighting methodology (and techniques for variance estimation) is detailed in Section \ref{weighting}.  In Section \ref{sec2}, we relay results from the application of these methods to the survey of military caregivers. Section \ref{sec5} presents simulation studies performed using data from the caregiver survey and purely synthetic data, and Section \ref{conclude} provides concluding points of discussion.  Supplementary materials consider extensions to settings involving multiple convenience samples, provide details not given in the main article, and tie together loose ends.

\section{Weighting Schemes} \label{weighting}

Prior to proposing weighting methods for blending probability and convenience samples, we introduce some fundamental concepts and notation that will be used throughout.  To simplify the exposition, we assume unstratified, unclustered designs (however, stratification and clustering are discussed in Section \ref{conclude}).  We let $\Omega$ denote the set of units (individuals) within a population, whereas $S_1 \subset \Omega$ and $S_2 \subset \Omega$ denote the set of units within the probability and convenience samples, respectively.  (In the supplemental materials, we extend theory outlined here to allow for multiple convenience samples.)  Under the assumption that the two samples are disjoint (i.e., $S_1 \cap S_2 = \emptyset$), we notate the set of respondents within the combined (or blended) sample as $S = S_1 \cup S_2$.  We assign a weight $w_i$ to each unit $i \in S$.  The problem of interest involves inferences for a variable $\{y_i\}_{i \in S}$, whereas a set of auxiliary variables (that does not include $y_i$), which are denoted by the vector ${\bf x}_i$ for unit $i$, are used to capture discrepancies between the two samples.  Although the theory presented here is applicable for multivariate outcomes, we assume univariate $y_i$ to simplify the exposition.  The notation $\{ \cdot \}_{i \in A}$ indicates a set of data values over a subset $A \subseteq \Omega$; when referencing all sampled units, the subscript $i \in S$ is excluded.  Both $y_i$ and ${\bf x}_i$ are observed for all $i \in S$.   For the remaining units of the population, neither case-level data values nor population totals are assumed to be known.

We also assume that $S_1$ has been subjected to an estimable non-response mechanism.  That is, $S_1 \subseteq S_1^*$, where $S_1^*$ is the set of individuals selected for participation within probability sampling.
Because the analyst has little control over the composition of the convenience sample, we do not consider non-response for $S_2$ (i.e., we embed non-response within the mechanism that generates $S_2$).

Note that in lieu of variances, we often discuss design effects.  The design effect, as proposed by \cite{kish65}, quantifies the inflation in variance that is a consequence of a weighting scheme.  Specifically, it represents the ratio of the variance of a weighted estimator with the variance of an equivalent unweighted estimator (i.e., under a simple random sampling design).

\subsection{Weights Based on Propensity Scores} \label{pssection}

If the probability of inclusion into a sample is known for each sampled unit and positive for all units in the population, one can use the reciprocal of such probabilities as sampling weights; these weights may be used to provide unbiased estimators of population quantities within sampling schemes that are not simple random sampling \citep{kish65}.
Generally, $P(i \in A)$ denotes the probability that individual $i$ is in set $A$; this expression is written equivalently as $P(A)$ for simplicity.  Our first set of procedures are based on the quantities,
\[
d_i = P(S_1 | {\bf x}_i),  ~~~~
q_i = P(S_2 | {\bf x}_i), ~~~~\mbox{and}~~~~
p_i = P(S | {\bf x}_i),
\]
which represent the probability of inclusion in the probability, convenience, and blended samples, respectively, for individual $i$ given the individual's set of auxiliary characteristics ${\bf x}_i$.

In the presence of a non-response mechanism, $\{d_i\}$ is derived as follows.  The practitioner assigns a sampling probability, denoted
\[
d_i^* = P(S_1^* | {\bf x}_i),
\]
to all $i \in \Omega$.  Next, $S_1$ is generated by the response mechanism
\[
r_i = P(S_1 | S_1^*, {\bf x}_i),
\]
and as a result $d_i = d_i^*r_i$.
For convenience, we write all probabilities listed above as being conditioned upon the same set of auxiliary variables ${\bf x}_i$.  Of course, this is not required.

Since $\{q_i\}$ and $\{p_i\}$ cannot be directly determined, and to help approximate them, we use estimated values of the probability that a sampled unit was included into the convenience sample, which is given by
\begin{equation} \label{gamma}
\gamma_i =P(S_2 | S_1 \cup S_2, {\bf x}_i).
\end{equation}
The quantity $\gamma_i$ is analogous to a propensity score as commonly used in treatment/control studies \citep{rosenbaum83} and to differentiate nonrespondents from respondents \citep{little02, bethlehem11}---therefore, weights that incorporate this quantity are referred to as propensity score-based weights.  Our approaches based on these weights mandate the following assumptions:

We assume the selection probabilities for the probability sample are based only on a set of design variables, ${\bf x}_i^*$, so that $d^*_i = P(S_1^*|{\bf x}_i^*)$ where ${\bf x}_i^* \subseteq {\bf x}_i$ and ${\bf x}_i^*$ is observed for all $i \in \Omega$.  Also, the response mechanism is dependent only on a set of variables, $\widetilde{\bf x}_i$, so that $r_i = P(S_1|S_1^*,\widetilde{\bf x}_i)$ where $\widetilde{\bf x}_i \subseteq {\bf x}_i$ and ${\bf x}_i^*$ is observed for all $i \in S^*_1$.

\renewcommand{\theassumption}{\arabic{assumption}}
\begin{assumption}
\label{assump1}
{\normalfont Design of $S_1^*$}: The selection probabilities for the probability sample depend only on a set of design variables ${\bf x}_i^*$, so that $d^*_i = P(S_1^*|{\bf x}_i) = P(S_1^*|{\bf x}_i^*)$ where ${\bf x}_i^* \subseteq {\bf x}_i$ and ${\bf x}_i^*$ is observed for all $i \in S_1^* \cup S_2$ with $d_i^*>0$ for all $i \in \Omega$.  Furthermore, $d_i^*$ is known for all $i \in S_1^* \cup S_2$.
\end{assumption}

\begin{assumption}
\label{assump1a}
{\normalfont MAR of $S_1^*$}: The response mechanism that yields $S_1$ is dependent only on a set of variables $\widetilde{\bf x}_i$, where $\widetilde{\bf x}_i \subseteq {\bf x}_i$ and $\widetilde{\bf x}_i$ is observed for all $i \in S^*_1 \cup S_2$.  That is, $r_i = P(S_1|S_1^*,{\bf x}_i)= P(S_1|S_1^*,\widetilde{\bf x}_i)$.  Furthermore, $r_i$ is expressed as an estimable function of $\widetilde{\bf x}_i$ with $r_i>0$ for each $i \in \Omega$.
\end{assumption}

It follows from Assumptions \ref{assump1} and \ref{assump1a} that $d_i>0$ for all $i \in \Omega$, and consequentially the probability sample covers the population.

Assumption \ref{assump1a} establishes that non-response in $S_1^*$ is missing at random, in the nomenclature of \cite{little02}, with respect to $\widetilde{\bf x}_i$.  That is, $P(S_1 | S_1^*,{\bf x}_i,z_i) = P(S_1 | S_1^*,{\bf x}_i)$, where $z_i$ is exogenous to $\widetilde{\bf x}_i$.  This observation and Assumption \ref{assump1} establish that the probability sample is conditionally independent the outcome of interest in that $P(S_1 | {\bf x}_i,y_i) = P(S_1 | {\bf x}_i)$.  We next impose a similar expression for the convenience sample.

\begin{assumption}
\label{assump2}
{\normalfont Ignorability:}
It holds that
\begin{equation} \label{mar}
P(S_2 | {\bf x}_i,y_i) = P(S_2 | {\bf x}_i).
\end{equation}
\end{assumption}

Letting $f( \: \cdot \: | \: \cdot \: )$ denote a conditional density, (\ref{mar}) can be equivalently stated as
\begin{equation*}
f(y_i | {\bf x}_i,S_2) = f(y_i | {\bf x}_i),
\end{equation*}
which is the exchangeability (or ignorability) assumption that underpins traditional propensity score techniques \citep{rosenbaum83}.

Note that the theory that forms the methodology presented here is valid in the event that the outcome $y_i$ is included in the set of auxiliary variables used for blending.  However, as illustrated in Section \ref{synthetic}, it is not beneficial to do so (specifically, no precision is gained by utilization of the convenience sample in such circumstances).
To ensure validity when $y_i \notin {\bf x}_i$, we impose Assumption \ref{assump2}.  To assess whether this assumption holds in practice, we suggest the test for the adequacy of blending as described in Section \ref{adequacy}.

\begin{assumption}
\label{assump3}
{\normalfont Model Adequacy:}
The models used to estimate $r_i$ and $\{\gamma_i\}$ are each correctly specified.
\end{assumption}

It is common practice in missing data analysis and in propensity score approaches to assume that the models used (i.e., non-response and propensity score models) are not misspecified---certainly, our procedures require the same.  In that vein, if the mechanism that generates the convenience sample is thought to be independent of some elements of ${\bf x}_i^*$ or $\widetilde{\bf x}_i$, the model for $\gamma_i$ should be adjusted accordingly.

\begin{assumption}
\label{assump4}
{\normalfont Positivity:}
For all $i \in \Omega$, $q_i > 0$.
\end{assumption}

Traditional propensity score methods require a positivity assumption (i.e., the propensity scores are non-zero).  Here, positivity is implied by assuming that the convenience sample covers the population.  However, as is clarified shortly, this assumption is not needed for all methods we present.

Robustness to these assumptions is assessed in Section \ref{sec5}.  Given the assumptions, we present two options for weighting via inclusion probabilities based on propensity scores; the first weights the samples separately, whereas the second jointly weights the samples to obtain representativeness.

\subsubsection{Disjoint Weighting}

Our first propensity score weighting scheme involves estimation of $\{q_i\}$, the probability of inclusion into the convenience sample.  Note that
\begin{equation*}
\gamma_i = \frac{P(S_2 | {\bf x}_i)}{P(S_1 \cup S_2 | {\bf x}_i)} = \frac{P(S_2 | {\bf x}_i)}{P(S_1 | {\bf x}_i)+P(S_2 | {\bf x}_i)} = \frac{q_i}{d_i + q_i},
\end{equation*}
where we have made use of the fact that $S_1 \cap S_2 = \emptyset$.  Solving the above expression for $q_i$ gives
\begin{equation} \label{qi}
q_i = \frac{d_i\gamma_i}{1- \gamma_i} = \frac{r_id_i^*\gamma_i}{1- \gamma_i}.
\end{equation}
Note that (\ref{qi}) does not indicate that $q_i$ is dependent upon $d_i$ (i.e., any alterations in $d_i$ while leaving $q_i$ unchanged will be offset by changes in the value of $\gamma_i$).

It follows from Assumptions \ref{assump1} and \ref{assump1a} that if weights calculated using $d_i^{-1}$ are applied to the probability sample, the probability sample will be representative of the population. Likewise, Assumption \ref{assump4} implies that weights calculated using $q_i^{-1}$, when applied to the convenience sample, will ensure that the convenience sample is representative.  That is,
\begin{equation} \label{disjoint}
\mbox{E}\left[\frac{\sum_{i \in S_1} d_i^{-1} y_i}{\sum_{i \in S_1} d_i^{-1}}\right]
\approx
\mbox{E}\left[\frac{\sum_{i \in S_2} q_i^{-1} y_i}{\sum_{i \in S_2} q_i^{-1}}\right]
\approx
\frac{\sum_{i \in \Omega} y_i}{N},
\end{equation}
Note that the first and second expectations in (\ref{disjoint}) are taken over the random variables used to generate $S_1$ and $S_2$, respectively.  The above result follows from the unbiasedness of the Horvitz-Thompson estimator and illustrates that representativeness for both panels can be obtained without combining the two samples.  Nonetheless, fusion of the data sources is preferred since blending is expected to yield estimators with lower variance than those which are constructed using the samples individually.

We next propose so-called {\bf disjoint propensity score} weights---the phrase ``disjoint'' blending is used since such procedures involve combination of two samples that are individually representative.  Analysis of multiple surveys using disjoint combing methods has been considered previously \citep[e.g.,][]{kott95, renssen97, merkouris04, lee06}.
Although the two samples are essentially weighted separately in disjoint blending, the weights are concatenated to yield a single set of weights that enables the two datasets to be analyzed in tandem.
Note that this mandates that the analyst decide how much emphasis to give one sample over the other in the concatenated weights.
Therefore, our first set of blending weights (notated $\{w^*_i\}$ for $i \in S$), is calculated by setting $w^*_{i}=\kappa d_i^{-1}$ for $i \in S_1$ and $w^*_{i}=(1-\kappa)q_i^{-1}$ for $i \in S_2$ for some constant $\kappa \in [0,1]$.

As illustrated in the supplemental materials to this article, $\{w^*_i\}$ may be used to obtain unbiased estimators of population quantities for any $\kappa \in [0,1]$.

In order to minimize the Kish approximation of the design effect (i.e., $\mbox{deff} = n\sum w_i^2 / (\sum w_i)^2$), we suggest
\[
\kappa = \frac
{(\sum_{i \in S_1} d_i^{-1}) (\sum_{i \in S_2} q_i^{-2})}
{(\sum_{i \in S_1} d_i^{-1}) (\sum_{i \in S_2} q_i^{-2})+(\sum_{i \in S_1} d_i^{-2})(\sum_{i \in S_2} q_i^{-1})}.
\]

We do not recommend selection of $\kappa$ so as to minimize the true design effect (or variance for that matter) since doing so would require a different value of $\kappa$ for each estimate produced.  However, such approaches have been considered previously \citep{hartley74, rivers03, lohr11}. Those authors suggest {\em post hoc} blending of estimators taken from multiple frames.  Therein, the two samples are combined after analyses involving outcomes begins.  (In our proposed methods, blending occurs prior to such analysis.) Consider the estimators $\hat\theta_1$ and $\hat\theta_2$ as described in the previous paragraph.
The {\em post hoc} blended estimator of $\theta$ is
\begin{equation} \label{posthoc}
\bar{\theta} = \bar{\kappa}\hat\theta_1 +(1-\bar{\kappa})\hat\theta_2
\end{equation}
where $\bar\kappa$ is selected to minimize the mean squared error of $\bar\theta$.  That is, set
\begin{equation} \label{barkap}
\bar\kappa = \frac{\mbox{Var}(\hat\theta_2)-\mbox{Cov}(\hat\theta_1,\hat\theta_2)}{\mbox{Var}(\hat\theta_1)+\mbox{Var}(\hat\theta_2)-2\mbox{Cov}(\hat\theta_1,\hat\theta_2)}.
\end{equation}
This calculation incorporates the assumed unbiasedness of $\hat\theta_1$ and $\hat\theta_2$ but does not assume $\theta_1$ and $\theta_2$ are independent (as $\theta_2$ is calculated using weights that invoke $S_1$).

Blending via disjoint weights as proposed with the use of the Kish approximation should yield a nearly minimum MSE estimator, and it does not require analyzation of the samples separately, nor does it require a different choice of $\kappa$ for each estimate.

\subsubsection{Simultaneous Weighting}
Weights may also be calculated using the probability of inclusion into the blended sample.  Note that $p_i = P(S_1 \cup S_2|{\bf x}_i) = P(S_1 |{\bf x}_i) + P(S_2|{\bf x}_i) = d_i + q_i $. Hence,
\begin{equation} \label{pi}
p_i = \frac{d_i}{1-\gamma_i} = \frac{r_id_i^*}{1-\gamma_i}
\end{equation}
for all $i \in S$.  Therefore, our second proposed choice of weights for blending probability and convenience samples is determined by setting $w_i=p_i^{-1}$ for all $i \in S$.  We refer $\{w_i\}$ as {\bf simultaneous propensity score} weights.
Note that Assumption \ref{assump4} is not required for simultaneous weighting.  That is, if $q_i = 0$, it follows that $\gamma_i = 0$ and thus $p_i = d_i >0$ (and $S$ covers $\Omega$).

Simultaneous blending, which has been described previously in \cite{lee09} in the context of calibration weighting, is designed to make the two samples individually representative of the population when used in tandem. However, when the samples are weighted in this manner, they are not individually representative.  That is, in general,
\begin{equation*}
\mbox{E}\left[\frac{\sum_{i \in S_1} w_i y_i}{\sum_{i \in S_1} w_i}\right] \neq \frac{\sum_{i \in \Omega} y_i}{N}
\end{equation*}
where $N$ is the cardinality of $\Omega$ (i.e., the population size).  Equivalence does not approximately hold in general in this expression.  Therefore, it is not prudent to use $\{w_i\}$ in conjunction with the observed data for the purpose of testing for the presence of discrepancies between the probability and convenience samples that are unexplained (which would determine whether or not $\{{\bf x}_i\}$ is sufficient for modeling inclusion probabilities).

Note that both disjoint and simultaneous propensity score methods require a known or estimated value of $d_i$ for each $i \in S_1 \cup S_2$.  In practice, these probabilities are easily determined.  From Assumption \ref{assump1}, $d_i^*$ is known for all $i \in \Omega$.  (E.g., if $n$ individuals from a stratum of size $N$ are randomly selected asked to participate in the study within probability sampling, one may use $d_i^* = n/N$ for all $i$ in the stratum's population.)  It is also assumed that the response mechanism, $r_i$, that yields $S_1$ is estimated as a function of predictor variables---for example, $\hat{r}_i = 1/(1+\exp\{-\hat\alpha_0 - \widehat{\bm{\alpha}}'\widetilde{\bf x}_i\})$.  Since $\hat{r}_i$ can be calculated for all $i \in S$, we approximate $d_i$ for $i \in S$ using $ d_i^*\hat{r}_i$.

\subsection{Calibration Weighting}

If population totals for a set of auxiliary variables ${\bf x}_i$, that is ${\bf t}_x = \sum_{i \in \Omega} {\bf x}_i$, are known, calibration weighting \citep{deville92, sarndal07} is useful for deriving weights that can be used to calculate generalizable estimators from non-representative data.
As such, it has a lengthy history as a tool for non-response adjustments \citep[e.g.][]{kott06b} and has a similar utility for blending probability and convenience samples.  Further, it should give results that are similar to those found using propensity score weighting (albeit through different computational procedures).

Calibration involves the determination of weights $\{v_i\}$ that satisfy
\begin{equation}\label{simulcali}
{\bf t}_x = \sum_{i \in S} v_i{\bf x}_i.
\end{equation}
Under a linear representation of $y_i$ as a function of ${\bf x}_i$ (e.g., $\mbox{E}[y_i] = \bm{\beta}'{\bf x}_i$), it follows that $\hat{t}_y = \sum_{i \in S} v_i {y}_i$ is a unbiased estimator of $t_y = \sum_{i \in \Omega} {y}_i$.  Similarly, unbiased estimates of $\mbox{E}[y_i]$ can be calculated using the weighted data.

Similar to the framework we established for weighting via propensity scores, we let weights $\{v_t\}$, which  satisfy (\ref{simulcali}) for $S=S_1 \cup S_2$, denote so-called { \bf simultaneous calibration} weights; these weights jointly blend the probability and convenience samples to produce a blended sample that is representative of the population.  However, it is also possible to individually weight the two samples so that each one is separately representative of the population.  That is, we may use calibration to find weights $\{v^*_{1i}\}_{i \in S_1}$ and $\{v^*_{2i}\}_{i \in S_2}$ that satisfy
\begin{equation}\label{disjointcali}
{\bf t}_x = \sum_{i \in S_1} v^*_{1i}{\bf x}_i
= \sum_{i \in S_2} v^*_{2i}{\bf x}_i.
\end{equation}
If we define weights $\{v^*_i\}_{i \in S}$ via $v^*_{i}=\kappa v^*_{1i}$ for $i \in S_1$ and $v^*_{i}=(1-\kappa) v^*_{2i}$ for $i \in S_2$ for some $\kappa \in [0,1]$, then $\{v^*_i\}$ is conducive for calculation of unbiased estimates of $\mbox{E}[y_i]$ and $\mbox{E}[y_i|{\bf x}_i]$.  We refer to $\{v^*_i\}$ as {\bf disjoint calibration} weights.  Similar to the disjoint propensity score-based weights, we set
\[
\kappa = \frac
{(\sum_{i \in S_1} v^*_{1i}) (\sum_{i \in S_2} (v^*_{2i})^2)}
{(\sum_{i \in S_1} (v^*_{1i})^2)(\sum_{i \in S_2} v^*_{2i})+(\sum_{i \in S_1} v^*_{1i}) (\sum_{i \in S_2} (v^*_{2i})^2)}
\]
in order to minimize the Kish approximation of the design effect.  As is the case with $\{w^*_i\}$ (the disjoint propensity score weights), $\{v^*_i\}$ can be used to test for post-weighting discrepancies between the two samples discussed in Section \ref{adequacy} below; the simultaneous calibration weights $\{v_i\}$ should not be used for this purpose.
Unlike the weights based on selection probabilities, calibrated weights may adjust for non-response within the probability sample (if the non-response mechanism is explained by the auxiliary variables ${\bf x}_i$).  Further, disjoint calibration weights can be used with {\em post hoc} blending estimation in the vein of (\ref{posthoc}).

Given initial design weights $\{\omega_i\}$, the calibration weighting procedure, denoted generalized raking by \cite{deville92}, calculates the set of calibrated weights $\{v_t\}$ which minimize the distance between the initial and calibrated weights subject to a distance metric $G(\cdot)$ and the constraints imposed by (\ref{simulcali}).  That is, the quantity $\sum_{i \in S}\omega_iG(v_i/\omega_i)$ is minimized subject to (\ref{simulcali}).  Calculations show that $\{\omega_i\}$ satisfies the expression $\sum_{i \in S}\omega_iF({\bf x}_i'\bm{\xi}){\bf x}_i = {\bf t}_x$, where $F(\cdot)$ is the inverse of the derivative of $G(\cdot)$ and where $\bm{\xi}$ is a vector of Lagrange multipliers (which may be extracted via Newton's method).  See \cite{deville93} for details.  We use a truncated linear distance metric bounded below at zero (i.e., $G(x)= (1/2)(x-1)^2$ for $x\geq 0$ and $G(x) = \infty$ otherwise), which has been shown to help minimize the design effect \citep{robbins17} while ensuring that $w_i > 0 $ for all $i \in S$.
The propensity score-based weights $\{w_i\}$ or $\{w^*_i\}$ can be used as initial values within calculation of their respective calibration weights $\{v_i\}$ or $\{v^*_i\}$.  However, it is also feasible to initialize the algorithm by giving each respondent equal weight.

Assumptions \ref{assump1}-\ref{assump2} are necessary when using calibration.  Note that unlike propensity scores, calibration may be applied without knowing $d_i$ for $i \in S_2$.
Since we do not use a propensity score model, Assumption \ref{assump3} is not needed.  Analogously, validity of calibration estimators is frequently illustrated on outcomes that obey $y_i = \beta_0 + \bm{\beta}'{\bf x}_i + \epsilon_i$ \cite[see][for example]{deville92}.  However, it commonplace to apply calibration estimation to outcomes that do not obey this formulation (e.g., binary data).  Assumption \ref{assump4} is needed for disjoint (but not simultaneous) calibration.

Calibration weighting for the purpose of blending probability and convenience samples has been proposed by \cite{disogra11}; therein, a technique akin to the simultaneous calibration method presented here is suggested.  \cite{lee09} propose a method akin to the disjoint calibration procedure discussed herein.

\subsubsection{Estimation of Benchmarks}

The vector of population totals ${\bf t}_x$ is said to be composed of benchmark values of the auxiliary variables.  It is unlikely that this vector will be known in practice, especially if the population of interest is a rare segment of a larger population.  In the event that no elements of ${\bf t}_x$ are known, one can approximate it using $\hat{\bf t}_x = \sum_{i \in S_1}d_i^{-1}{\bf x}_i$. where $\{d_i^{-1}\}$ are the design weights.  Calibration weights calculated using $\hat{\bf t}_x$ offer no perceived advantage over the propensity score-based weights.  Nonetheless, calibration weighting can be used to incorporate limited aggregate information regarding the population of interest.  As an example: If ${\bf x}_i$ can be decomposed into two vectors ${\bf x}_i^{(1)}$ and ${\bf x}_i^{(2)}$, satisfying ${\bf x}_i=(({\bf x}_i^{(1)})',({\bf x}_i^{(2)})')'$, where ${\bf t}_x^{(1)} = \sum_{i \in \Omega} {\bf x}_i^{(1)}$ is known but ${\bf t}_x^{(2)} = \sum_{i \in \Omega} {\bf x}_i^{(2)}$ is not, one can apply calibration to find weights $\{v_i^{(1)}\}$ that obey $\sum_{i \in S_1} v_i^{(1)}{\bf x}_i^{(1)} = {\bf t}^{(1)}_x$.  Setting $\tilde{\bf t}_x = \sum_{i \in S_1} v_i^{(1)}{\bf x}_i$, the vector $\tilde{\bf t}_x$ yields benchmark values that can be used in calculation of blending weights $\{v_i\}$, $\{v_{1i}^*\}$ and $\{v_{2i}^*\}$.  Calibration weighting in this manner may account for non-response among the probability sample (so long as the non-response mechanism is influenced by only variables within ${\bf x}_i^{(1)}$).

\subsection{A Test for the Adequacy of Blending} \label{adequacy}

We are interested in knowing whether or not the blended sample is representative of the population (following application of the weighting methods described above).  If the probability of inclusion in the convenience sample has not been appropriately modeled or if an insufficient scope of auxiliary variables has been used within calibration weighting, the blended sample is not assured to be representative.  Therefore, we develop a method for statistically testing the adequacy of blending.

Let ${\theta}$ denote a population parameter of interest.  Further, let $\hat\theta_1$ and $\hat\theta_2$ denote estimates of $\theta$ found using only $i \in S_1$ and only $i \in S_2$, respectively.  We will test to see whether or not $\hat\theta_1$ and $\hat\theta_2$ are statistically equivalent.  Lack of equivalence may indicate the presence of a latent characteristic that exists outside the set of auxiliary variables $\{{\bf x}_i\}$ which differentiates $S_1$ and $S_2$ (in which case (\ref{mar}) is violated).
If $\hat\theta_1$ and $\hat\theta_2$ are calculated using simultaneous blending weights, we expect a non-negligible discrepancy between $\hat\theta_1$ and $\hat\theta_2$ when blending is adequate (even though $\hat\theta$, an estimator of $\theta$ determined using data from both samples with simultaneous blending weights, is expected to be unbiased).  To observe $\hat\theta_1 \approx \hat\theta_2$ under an adequate selection of ${\bf x}_i$, disjoint blending weights (determined using propensity scores or calibration) are needed.

We propose a test for the adequacy of blending that is based off of the mean value of an outcome of interest $\{y_i\}$.  Letting $\mu = \sum_{i \in \Omega}y_i/N$, we evaluate differences between $\hat\mu_1$ and $\hat\mu_2$, which are estimates of $\mu$ based off of the probability and convenience samples, respectively, and which are calculated using a disjoint weighting scheme.  We approach this problem by fitting the model
\begin{equation} \label{deltaeq}
y_i = \mu + \delta 1_{\{i \in S_2\}} + \epsilon_i,
\end{equation}
for each $i \in S$ using weighted least squares.  We let $\{\epsilon_i\}$ denote mean zero errors and, generally, $1_{\{A\}}$ represents the indicator of event $A$.  We test
\begin{equation} \label{h0}
{\cal H}_0: \delta = 0 ~~~\mbox{against}~~~ {\cal H}_1: \delta \neq 0.
\end{equation}
This test may be performed using the statistic $z^* = \hat\delta/\widehat{\mbox{Var}}(\hat\delta)^{1/2}$.

We formulate the test using the linear model in (\ref{deltaeq}) as opposed to a basic two-sample formulation so as to enable easy extension to analogous tests that involve non-linear response (e.g., logistic regression) or that involve regression models with additional covariates.
One may also develop an omnibus version of the above which can incorporate multiple outcomes (although this is not considered further here).

\subsection{Variance Estimation} \label{var}

We briefly discuss techniques for variance estimation in the presence of weights calculated via the methodology outlined herein.
We first describe (Taylor series) linearization and conclude with a discussion of resampling methods (with a focus on the jackknife).
Further details on both techniques are given in the supplementary materials to this article.

Linearization provides algebraic approximations of variances of estimators found using complex survey data.  The procedure invokes Taylor series expansions to translate an estimator into terms for which the variance can be estimated using straightforward procedures designed for complex surveys \citep[e.g., the Horvitz-Thompson estimator][]{horvitz52}.  We implement linearization through the \texttt{survey} package in \texttt{R} \citep{lumley04, lumley11}.

By using resampling techniques, one can incorporate aspects of an estimation process into variance calculations that are not easily captured algebraically. We focus on the jackknife \citep{quenouille49, quenouille56}, wherein data are segmented into replication groups, and the estimation process is applied to each replication group separately. Here, we consider a delete-a-group jackknife \citep{kott01}, wherein samples are segmented into $G$ mutually exclusive and exhaustive groups, and the weights are recalculated for each group.

\section{A Survey of Military Caregivers} \label{sec2}

The early 21$^{\rm st}$ century wars in Afghanistan and Iraq (Operation Enduring Freedom, Operation Iraqi Freedom, and Operation New Dawn) have produced a wave of returning wounded veterans and servicemembers.  Many of those who have engaged in these modern conflicts have suffered unique traumas (e.g., traumatic brain injury associated with improvised explosive devices); fortunately, they have also benefited from advances in battlefield medicine and rehabilitative services (thereby reducing death rates overall but resulting in an increased proportion of veterans who have endured disabling injuries).  As a result, America's modern-day wounded warriors are often in need of a variety of caregiving services (e.g., bathing, dressing, and eating, etc.) upon their return to the homefront.  However, these caregiving burdens often fall upon a range of family members, friends and other acquaintances (i.e., military caregivers) who do not receive wages for their services.  As recently as a decade following the beginning of the conflicts in Iraq and Afghanistan, little was known about the average American military caregiver (e.g., their prevalence across the American population, the specific obstacles they face, etc.); this lack of information acted as an impediment for policymakers aiming to establish services to assist caregivers.

\cite{ramchand14} report findings from a large study of military caregivers performed from August to October 2013 by the RAND Corporation and sponsored by the Elizabeth Dole Foundation---our empirical application here utilizes data from this study.
Substantive definitions utilized within \cite{ramchand14} are reviewed here.  Generally, a caregiver is someone (a family member, friend, neighbor, etc.) who provides a broad range of care for an individual with a disabling (physical or mental) wound, injury or illness.  A key characteristic of caregivers in our context is that they do not receive wages for caregiving.  A military caregiver provides these services to a current or former member of the U.S.~Armed Services.  Further, a military caregiver is denoted as post-9/11 if the care recipient served in the armed forces at any point following September 11, 2001 (regardless of prior military service).

A crucial aspect of the RAND survey of military caregivers is that it was probability-based. A key policy question investigated by the military caregivers study involved the comparison of caregivers of servicemembers who served during the Afghanistan and Iraq conflicts (i.e., post-9/11 caregivers) to caregivers of servicemembers who served in prior eras (i.e., pre-9/11 caregivers).  It was understood that probability-based sampling would not yield a sufficient number of post-9/11 caregivers; therefore, convenience samples from this segment were also drawn.  Here, we apply the methodology described in Section \ref{weighting} in order to blend data from the probability-based sample of caregivers with data from a convenience sample of post-9/11 caregivers and thus enable inferences that are generalizable to this sub-population.

\cite{ramchand14} describe a multitude of analyses that were performed using the caregiver data.  Primarily, marginal means for respondents within the domains described above are calculated for several outcomes of interest.  A primary outcome discussed here is the depression score, which is gauged using the eight-question Patient Health Questionnaire \citep{kroenke09}.  A related outcome measure discussed here is anxiety levels, which are measured using the Mental Health Inventory anxiety subscale.  On these two scales, higher scores indicate higher levels of depression and anxiety.

\subsection{The Probability Sample: KnowledgePanel (KP)}

The GfK KnowledgePanel \citep[KP;][]{gfk13}, which is a nationally representative panel of approximately 45,000 American adults (during August 2013), was used as a probability sample for the military caregivers survey.  Each member of the KnowledgePanel was given a screener and was consequentially placed into one of the following domains: 1) military veteran receiving unpaid care, 2) military caregiver, 3) civilian caregiver, and 4) other (i.e., non-veteran/non-caregiver).  All respondents who fell into the first two domains listed were given a full survey.  Caregivers were asked to report on themselves and their care recipient; likewise, veterans were asked to report on themselves and their caregiver.  Random samples of members who fell into the final two domains were selected and asked to complete a full survey.  Of the 41,163 panelists invited to complete the screener, 28,164 (68\%) did so.

To facilitate blending of data from supplemental sources, military caregivers who completed the full survey are then further segmented on the basis of the criteria mentioned earlier.  Table \ref{nbystratum} lists the number of respondents from each domain that were sampled from the KP (as well as the respondents from the convenience samples, which are described later).

\begin{table}[!ht]
\centering
\small
 \caption{The number of respondents by domain and panel.
}   \label{nbystratum}
\begin{tabular}{lcccc}
\hline
Domain		&	KP	&	WWP	\\
\hline
Veterans		&	251	&	--	\\
Post-9/11 Caregiver&	72	&	281	\\
Pre-9/11 Caregiver	&	522	&	3	\\
Civilian Caregiver	&	1828	&	--	\\
Non-Caregiver	&	1163	&	--	\\
\hline
 \end{tabular}
\end{table}
Pre-stratification weights (i.e., screener weights) for all screener respondents from the KP were calculated by using calibration with a set of standard demographic variables (e.g., age, gender, income, etc.) as auxiliary variables (see ${\bf x}_i$ in (\ref{simulcali})) where the benchmark values (${\bf t}_x$) are determined using information (or the non-institutionalized U.S.~adult population) from the 2010 Census and American Community Survey \citep[see][for futher details]{ramchand14}.  These weights account for non-response at the screener stage.  In the ensuing, military caregivers are segmented further by era of service (pre- vs.~post-9/11), and we focus on the domain of post-9/11 caregivers.

\subsection{The Convenience Sample: Wounded Warrior Project (WWP)}\label{sec3}

As indicated by Table \ref{nbystratum}, there are only 72 post-9/11 caregivers from the probability sample, which yields concern that efforts to compare this domain to other domains will be under-powered.  To further supplement our data, we used a database from the Wounded Warrior Project (WWP).
The WWP is a nonprofit veterans service organization that offers a variety of programs to support servicemembers who were wounded during the military actions following the events of September 11, 2001.
The WWP maintains a database of individuals who have registered with the organization as caregivers of wounded, ill, or injured veterans of post-9/11 military actions---this database yielded a sampling frame for additional post-9/11 caregivers.
The rate of response among individuals sampled from the WWP was 20\%, which yielded 281 post-9/11 caregivers who completed the survey.

We compare the KP and WWP across a wide range of variables.  Tables \ref{bigtable} and \ref{bigtable1} show (among other information) means and standard errors of several variables from the caregiver survey for respondents from the KP and the WWP.   We note that Table \ref{bigtable} includes auxiliary variables used in weighting, whereas Table \ref{bigtable1} includes a selection of outcome variables (although this discrepancy is not yet germane to the discussion).  Since the WWP contains primarily post-9/11 caregivers, the information in these tables is calculated using only data from post-9/11 caregivers.  The table also lists {\em unweighted} means and standard errors of these variables when calculated using the KP and WWP jointly.  Note that the standard errors presented in these tables (and in all subsequent analyses) are calculated using Taylor series linearization.  As indicated by Tables \ref{bigtable}, there are substantial differences across the KP and WWP panels.  Specifically, the WWP caregivers are predominantly female (94\%), whereas around half of the caregivers from the KP are female.  Likewise, around 69\% of the caregivers from the WWP are caring for a veteran with traumatic brain injury (TBI), whereas only 16\% of care recipients from the KP have TBI.  Furthermore, WWP caregivers report higher depression and anxiety levels.  Note that all differences in means seen in Table \ref{bigtable} are statistically significant at the 5\% level.
\begin{table}[!ht]
\centering
\begin{threeparttable}
\caption{
Means and standard errors of selected survey variables for post-9/11 caregivers calculated using various sources of data.
} \label{bigtable}
{\small
\renewcommand{\tabcolsep}{.15cm}
\begin{tabular}{lcccccccc}
\hline
 &   \multicolumn{2}{c}{KP Only} &  &  \multicolumn{2}{c}{WWP Only} &  & \multicolumn{2}{c}{\specialcell{Blended: \\ Unweighted}}  \\  \cline{2-3} \cline{5-6} \cline{8-9}
Variable Description &   Mean & s.e. &  & Mean & s.e. &  & Mean & s.e. \\
\hline
Caregiver lives with veteran &   0.453 & 0.085 &  & 0.858 & 0.021 &  & 0.755 & 0.032 \\
Veteran deployed to war zone &   0.577 & 0.086 &  & 0.929 & 0.015 &  & 0.839 & 0.029 \\
Vet.~has a service-related disability rating &   0.563 & 0.095 &  & 0.883 & 0.016 &  & 0.801 & 0.031 \\
Vet.'s disability rating is 70+\% &   0.280 & 0.086 &  & 0.715 & 0.023 &  & 0.605 & 0.033 \\
Vet.~has serv.-rel.~traumatic brain injury &   0.164 & 0.069 &  & 0.687 & 0.027 &  & 0.554 & 0.032 \\
Vet.~has serv.-rel.~mental health problems &   0.450 & 0.097 &  & 0.897 & 0.021 &  & 0.783 & 0.033 \\
Caregiver is female &   0.476 & 0.082 &  & 0.940 & 0.014 &  & 0.822 & 0.032 \\
Caregiver depression level\tnote{\dag} &   7.071 & 0.808 &  & 9.485 & 0.446 &  & 8.871 & 0.356 \\
Caregiver anxiety level\tnote{\dag} &   38.409 & 5.221 &  & 50.896 & 1.727 &  & 47.722 & 3.887 \\
Caregiving caused quitting work entirely &   0.204 & 0.065 &  & 0.502 & 0.028 &  & 0.426 & 0.031 \\
Caregiving disturbs sleep &   0.474 & 0.090 &  & 0.808 & 0.028 &  & 0.723 & 0.030 \\
Caregiving causes financial strain &   0.588 & 0.085 &  & 0.762 & 0.021 &  & 0.718 & 0.028 \\
Caregiver feels overwhelmed by caregiving &   0.398 & 0.085 &  & 0.815 & 0.024 &  & 0.709 & 0.032 \\
\hline
\end{tabular}
}
\begin{tablenotes}
\item[\dag] Variable is ordinal or continuous---variables are binary otherwise.
\end{tablenotes}
\end{threeparttable}
\end{table}

\subsection{The Blending Weights} \label{sec4}

Since the unweighted respondents from the WWP report markedly different characteristics than those given by respondents from the probability sample (KP) and since policymakers need nationally representative estimates, we apply the weighting techniques of Section \ref{weighting} for the purpose of blending KP and WWP samples.

\subsubsection{Calculation of Propensity Scores} \label{calcprop}

As outlined in Section \ref{weighting}, calculation of propensity score-based weights requires two sets of probabilities as input:~$\{d_i\}$ and $\{\gamma_i\}$.  Recall that it is required that $d_i$, which gives the likelihood of unit $i$ being selected into the probability sample, be known for each $i \in S$.
Recall also that a pre-stratification weight was assigned to each post-9/11 caregiver within the KP and that these weights incorporate adjustments for non-response among the probability sample (i.e., non-compliance with the initial KP screener) and any residual inconsistencies between the KP and the full population; therefore, we set $\{d_i\}_{i \in S_1}$ equal to the inverse of the initial pre-stratification weights.
Due to the proprietary nature of the KP, we lack data on screener non-respondents, and therefore cannot fully model
screener non-response so as to calculate an informed value of $\{d_i\}_{i \in S_2}$. Hence, we set $d_i = n_1/\sum_{j \in S_1}d_j^{-1}$ for each $i \in S_2$, where $n_1$ is the sample size of $S_1$---this effectively assumes that all cases in $S_2$ had equal probability of inclusion into $S_1$.  For non-response adjustments that are more rigorous, we rely on calibration weights.

Next, we calculate $\gamma_i$, the probability unit $i$ sampled unit being selected within the convenience sample for each $i \in S$ (i.e., the propensity score).  These probabilities are unknown but can be easily estimated.  Specifically, we use logistic regression:
\begin{equation} \label{gammamod}
\log\left(\frac{\gamma_i}{1-\gamma_i}\right) = \zeta_0 +\bm{\zeta}'{\bf x}_i,
\end{equation}
where ${\bf x}_i$ is a set of auxiliary variables (the same set is used within calibration weighting) and $(\zeta_0,\bm{\zeta}')'$ is a vector of regression parameters.  As a substitute for $\{\gamma_i\}$, we use $\{\hat\gamma_i\}$, which contains the predicted values (derived via the above regression) of the elements of $\{\gamma_i\}$. Non- and semi-parametric models for binary response could be used \citep[e.g.,][]{hahn98, hirano03, frolich06}, and boosted regression models \citep{friedman01, friedman02, ridgeway14} have become increasingly popular for propensity score estimation; however, such techniques are not given further consideration here.

\subsubsection{Calibration Weighting}

When applying calibration here and in Section \ref{sec5}, we use the function \texttt{calibrate()} within the \texttt{survey} package in \texttt{R} \citep{lumley04, lumley11}.  Specifically, we set \texttt{calfun = 'linear'} and \texttt{bounds = c(0,Inf)}. As initial weights in the simultaneous calibration method, weights from the propensity score-based scheme were used.  Details regarding calculation of benchmarks for calibration in this application are provided in the supplemental materials.

\subsubsection{Results from Blending}

We calculated three sets of weights to blend the samples of post-9/11 caregivers: 1) simultaneous propensity scores (SPS), 2) disjoint propensity scores (DPS) and 3) simultaneous calibration (SC).  Weights based on disjoint calibration could not be calculated since the algorithm to do so failed (this is likely due to KP and WWP respondents being substantially different and thereby disabling the feasibility of a solution to the calibration equations in (\ref{disjointcali})).
Standard errors reported in this section were calculated using a delete-a-group jackknife with $G=40$ and with reweighting for each replicate group.

A subset of auxiliary variables (${\bf x}_i$) used in calculation of all weights includes all variables listed in Table \ref{bigtablea}.  These variables include five questions that gauge the status of the caregiver as being an ``early adopter'' of technology.  Such questions have proven to be useful for weighting respondents of online surveys \citep{disogra11} by helping to quantify their familiarity with and use of electronic media.  Table \ref{bigtablea} also lists the estimated mean of each variable when calculated using each of the three sets of blending weights.  Standard errors of these auxiliary variables are not provided. Note that weight trimming \citep{lee11, potter15} was used to reduce the influence of outlying weights.  Specifically, as the final step in weighting, the highest and lowest 1\% of weights were truncated to match the corresponding percentile, and weights within these bounds were adjusted so that the sum of the weights remains unchanged.  As a consequence, the mean of the auxiliary variables when calculated using calibration weighting differs slightly from the benchmarks values.  Additionally, Table \ref{bigtablea} indicates that weights based on propensity scores yield means of calibration variables that closely approximate the benchmark values.  The full set of auxiliary variables for blending is shown in Table A.2 of the supplemental materials along with information analogous to that which is provided in Tables \ref{bigtable} and \ref{bigtablea}.

\begin{table}[!ht]
\centering
\begin{threeparttable}
\caption{
Benchmarks and weighted averages among post-9/11 caregivers for selected {\em auxiliary} variables (${\bf x}_i$).  Three schemes for calculating blending weights are shown:~simultaneous propensity scores (SPS), disjoint propensity scores (DPS), and simultaneous calibration (SC).
} \label{bigtablea}
{\small
\renewcommand{\tabcolsep}{.15cm}
\begin{tabular}{lccccc}
\hline
  & Bench- &  & \multicolumn{3}{c}{Blended: Weighted } \\  \cline{4-6}
Variable Description
 & marks &  & SPS & DPS & SC\\  \hline
Caregiver lives with veteran
 & 0.453 &  & 0.467 & 0.573 & 0.489 \\
Veteran deployed to war zone\tnote{\ddag}  & 0.559 &  & 0.592 & 0.631 & 0.591 \\
Veteran has a service-related disability rating
 & 0.563 &  & 0.577 & 0.648 & 0.591 \\
Veteran's disability rating is 70+\%
 & 0.280 &  & 0.275 & 0.320 & 0.308 \\
Vet.~has service-related traumatic brain injury
 & 0.164 &  & 0.133 & 0.166 & 0.189 \\
Vet.~has service-related mental health problems\tnote{\ddag}
 & 0.510 &  & 0.491 & 0.529 & 0.549 \\
Caregiver is female\tnote{\ddag}
 & 0.555 &  & 0.509 & 0.557 & 0.592 \\
\hline
\end{tabular}
}
\begin{tablenotes}
\item[\ddag] Veterans' reports are used in calculation of the benchmark.
\end{tablenotes}
\end{threeparttable}
\end{table}

Our set of auxiliary variables chosen not only because the KP and WWP caregivers differ on the basis of these characteristics but also because the selection of auxiliary variables was limited to demographic characteristics of the caregiver and descriptors of the care recipient.
Descriptors of caregiver well-being are outcome variables for this study.  Table \ref{bigtable1} lists selected outcome variables when calculated using only the KP caregivers, only the WWP caregivers, and while using the three sets of blended weights.
Table A.3 provides this information for a larger set of outcome variables.
For each method of blending (e.g., SPS, DPS, SC) and for each outcome variable, a $p$-value of the hypotheses in (\ref{h0}) is provided.  Corresponding $p$-values calculated using unweighted data (not shown for brevity) quantify the discrepancies between the two samples.
Specifically, of the 31 outcomes listed in Table A.3, 23 have an unweighted $p$-value less than 0.05 (16 have a $p$-value less than 0.01).

\begin{table}[!ht]
\centering
\begin{threeparttable}
\caption{Means and standard errors of selected {\em outcome} variables for post-9/11 caregivers from blended data with various weighting schemes.
The $p$-values are derived from a test of the hypotheses in (\ref{h0}).  The weighting schemes shown are simultaneous propensity scores (SPS), disjoint propensity scores (DPS), and simultaneous calibration (SC).} \label{bigtable1}
{\footnotesize
\renewcommand{\tabcolsep}{.12cm}
\begin{tabular}{lccccccccccc}
\hline
&  \multicolumn{3}{c}{Blended: SPS} &  & \multicolumn{3}{c}{Blended: DPS} &  & \multicolumn{3}{c}{Blended: SC}  \\  \cline{2-4} \cline{6-8} \cline{10-12}
Variable Description &  Mean & s.e. & $p$-val. &  & Mean & s.e. & $p$-val. &  & Mean & s.e. & $p$-val. \\  \hline
Caregiver depression level\tnote{\dag} &  7.661 & 0.575 & 0.109 &  & 7.574 & 0.895 & 0.513 &  & 7.801 & 0.668 & 0.112 \\
Caregiver anxiety level\tnote{\dag} &   41.551 & 3.887 & 0.084 &  & 43.657 & 4.827 & 0.344 &  & 41.706 & 3.933 & 0.340 \\
Caregiving caused quitting work &   0.206 & 0.072 & 0.009 &  & 0.237 & 0.066 & 0.489 &  & 0.237 & 0.069 & 0.114 \\
Caregiving disturbs sleep &  0.499 & 0.069 & 0.005 &  & 0.536 & 0.089 & 0.298 &  & 0.554 & 0.068 & 0.061 \\
Caregiving causes financial strain  & 0.593 & 0.064 & 0.174 &  & 0.627 & 0.089 & 0.557 &  & 0.575 & 0.067 & 0.327 \\
Feels overwhelmed by caregiving  & 0.443 & 0.104 & 0.007 &  & 0.480 & 0.078 & 0.291 &  & 0.480 & 0.087 & 0.010 \\
\hline
\end{tabular}
}
\begin{tablenotes}
\item[\dag] Variable is ordinal or continuous---unless otherwise specified, variables are binary.
\end{tablenotes}
\end{threeparttable}
\end{table}

As seen by comparing of Tables \ref{bigtable} and \ref{bigtable1} (and by examining Table A.3, the means of the outcome variables when found using weighted (blended) data tend to be close to the means found using only the KP sample; uniformly, the blended means are well within the error bound for the values found from the KP.  Additionally, the standard errors of the blended values imply that an improvement in precision has been obtained by using the convenience sample.  However, we see that simultaneous blending offers smaller standard errors (and therefore a greater increase in precision) than disjoint blending.  This observations yield the conclusion that the simultaneous blending weights are likely to provide smaller design effects---this claim is investigated further in Section \ref{sec5}.
In some circumstances, disjoint blending yields larger standard errors than were observed when only the KP was used.  This is a consequence of the blended data having a higher mean than was observed in only the probability sample, and is not evidence of a loss of precision.

The $p$-values reported in Table \ref{bigtable1} (and Table A.3) are often close to zero when simultaneous weights are used---this is expected and is not a testament to the quality of weighting.  For the disjoint blending method, the $p$-values are larger---these $p$-values give a more accurate assessment of the adequacy of weighting (with respect to the selection of auxiliary variables).  None of the 31 outcomes in Table A.3 have a $p$-value less than 0.05 under disjoint blending---this provides evidence that the set of auxiliary variables in Table \ref{bigtable} sufficiently account for differences between KP and WWP post-9/11 caregivers.

When disjoint weights were used, we also calculated the {\em post hoc} blending estimator described in (\ref{posthoc}); however, full results are omitted for brevity.  We note that the point estimates yielded by the {\em post hoc} method are consistently similar to those seen for the DPS procedure in Table \ref{bigtable1}.  For example, in the depression and anxiety outcomes, the {\em post hoc} estimators are 7.30 and 42.97, respectively (compared to respective values of 7.57 and 43.66 for DPS).  The {\em post hoc} blending estimator is examined further in Section \ref{sec5}.

As a final data analysis, we discuss the influence of weighting on inferences drawn from the following regression model of a military caregiver's depression level:
\begin{equation} \label{depmod}
\mbox{DEP}_i = \eta_0 + \eta_1 \mbox{ERA}_i + \eta _2 \mbox{AGE}_i + \eta_3 \mbox{SEX}_i + \eta_4 \mbox{INC}_i + \eta_5 \mbox{EMP}_i + \epsilon_i.
\end{equation}
In the above, $\mbox{DEP}_i$ is the depression score for the $i^{\rm th}$ caregiver, $\mbox{ERA}_i$ is an indicator variable which is unity if the caregiver is post-9/11, and $\mbox{AGE}_i$, $\mbox{SEX}_i$, $\mbox{INC}_i$ and $\mbox{EMP}_i$ are the age (continuous), gender (binary), income (continuous) and employment status (binary), respectively, of the caregiver.
Using all military caregivers (pre- and post-9/11) from the KP and WWP, we estimate the model in (\ref{depmod}).  The various blending weights are used for post-9/11 caregivers, and the pre-stratification weights are applied to the pre-9/11 caregivers; results are seen in Table \ref{depmodtab}.

\begin{table}[!ht]
\centering
\small
\begin{tabular}{clccccccc}
\hline \hline
 & & (Intercept) & ERA & AGE & SEX & INC & EMP \\ \hline
\multirow{3}{*}{KP Only} & Estimate & 11.4897 & 1.5102 & -0.7542 & -0.7812 & -1.4843 & -0.6085 \\
 & Std. Error & 1.8073 & 0.9375 & 0.3330 & 0.6134 & 0.4848 & 0.7070 \\
 & $p$-value & 0.0000 & 0.1078 & 0.0239 & 0.2034 & 0.0023 & 0.3898 \\  \hline
\multirow{3}{*}{\specialcell{Blended: \\ Unweighted}} & Estimate & 12.2617 & 2.3185 & -0.8561 & -1.6111 & -1.6121 & -0.1945 \\
 & Std. Errors & 1.3262 & 0.5632 & 0.2727 & 0.4990 & 0.3540 & 0.5224 \\
 & $p$-value & 0.0000 & 0.0000 & 0.0017 & 0.0013 & 0.0000 & 0.7098 \\  \hline
\multirow{3}{*}{\specialcell{Blended: \\ SPS}} & Estimate & 11.1934 & 1.6835 & -0.7676 & -1.1130 & -1.3655 & -0.2015 \\
 & Std. Error & 2.6438 & 0.9258 & 0.4008 & 0.7327 & 0.8287 & 1.1329 \\
 & $p$-value & 0.0002 & 0.0778 & 0.0639 & 0.1380 & 0.1086 & 0.8599 \\  \hline
\multirow{3}{*}{\specialcell{Blended: \\ DPS}} & Estimate & 11.5365 & 1.9340 & -0.7633 & -1.2863 & -1.4557 & -0.3199 \\
 & Std. Error & 2.6331 & 0.6637 & 0.3003 & 0.4554 & 0.6961 & 1.1046 \\
 & $p$-value & 0.0001 & 0.0063 & 0.0158 & 0.0079 & 0.0440 & 0.7739 \\  \hline
\multirow{3}{*}{\specialcell{Blended: \\ SC}} & Estimate & 11.6872 & 2.1431 & -0.8125 & -1.3930 & -1.4348 & -0.3934 \\
 & Std. Error & 1.8824 & 0.7785 & 0.3195 & 0.5647 & 0.7694 & 0.6950 \\
 & $p$-value & 0.0000 & 0.0094 & 0.0157 & 0.0188 & 0.0709 & 0.5751 \\  \hline
 \end{tabular}
 \caption{Results for the models in (\ref{depmod}) for various weighting schemes.  See the caption to Table \ref{bigtable} for a description of the acronyms.  }
  \label{depmodtab}
\end{table}

Of primary interest is whether or not $\eta_1$ differs from zero; if $\eta_1 > 0$, then post-9/11 caregivers have higher depression levels (conditional on covariates) than pre-9/11 caregivers.  Using weighted least squares (WLS) with the initial pre-stratification weights, the regression model is estimated while incorporating data from only military caregivers from the KP sample.  When only military caregivers from the probability sample are used, the results illustrate that post-9/11 caregivers have higher levels of depression, but the estimate is not statistically different from zero at the 5\% significance level.  The lack of statistical significance is likely a consequence of sample size.  All methods of calculating blending weights implicate that era of service is statistically significant (at the 5\% level) with post-9/11 caregivers having higher levels of depression when controlling for other characteristics.  However, the era of service coefficient  is less statistically significant when calculated using disjoint blending.  Further, the discrepancy between levels of depression for pre- and post-9/11 caregivers is smaller when any method of blending is used than when the unweighted WWP data are used.   The emotional hardships endured by post-9/11 military caregivers are thought to be a consequence (in part) of the types of injuries suffered by soldiers that served in the recent campaigns and the fact that these caregivers are often spouses and family members tending to a young veteran with a debilitating injury.

The sample means and standard errors reported in this section are calculated using the \texttt{svymean()} function from the \texttt{survey} package in R.  Regression results and $p$-values for comparing samples are derived using the \texttt{svyglm()} function.

\section{Simulations} \label{sec5}

We perform simulation studies to examine efficacy of the weighting methodology in greater detail.  First, we provide simulations performed using data from the caregiver study that are designed to assess bias and mean-squared error in situations where model assumptions are and are not met.  We also perform simulations with purely synthetic data that are designed to illustrate the efficacy of variance estimators.

\subsection{Simulations with caregiver data} \label{real}

Our first simulation study uses data from the caregiver survey.  We form a pseudo-population of post-9/11 caregivers using the observed data and then repeatedly draw samples from this population.   Prior to describing construction of the pseudo-population, we note that an opt-in sample military caregivers was drawn from a volunteer internet-based panel---this sample yielded 171 post-9/11 caregivers (see Table \ref{nbystratum}).  The opt-in sample was found to contain data anomalies (e.g., clearly erroneous response patterns) and was disregarded from the analyses provided by \cite{ramchand14}.  However, it is used here to help build the pseudo-caregiver population.  Specifically, the pseudo-population of post-9/11 caregivers consists of all observed military caregivers from the KP and opt-in samples.  The WWP is excluded from this phase of our study because it represents a highly differentiated segment of the caregiver population (e.g., the WWP is nearly entirely post-9/11 and is heavily burdened).  Our pseudo-population has 940 cases.

Approximately 150 caregiver are selected for inclusion in the probability sample by drawing units from the pseudo-population at random with probability $d_i^*=0.16$.  Selected caregivers are considered respondents with a probability $r_i$ that satisfies $\log(r_i/(1-r_i))= (1/3)\mbox{F}_i - (2/3)\mbox{A}_i$, where $\mbox{F}_i$ is an indicator of whether or not the caregiver is female and $\mbox{A}_i$ is the caregiver's age categorized in a 5-point Likert scale (both variables are centered).
A convenience sample of is drawn from the pseudo-population by using the following logistic model:
\begin{equation} \label{convmod}
\log\left(\frac{\rho_i}{1-\rho_i}\right) = b_0 + \bm{b}_1'{\bf v}_i,
\end{equation}
where $\rho_i$ is the probability that case $i$ is assigned to the convenience sample.  Further, ${\bf v}_i$ is a set of variables upon which the probability and convenience samples may be differentiaed and $(b_0,\bm{b}_1')'$ is a vector of coefficients.  When applying (\ref{convmod}), we use standardized versions of all variables so as to ensure that the coefficients $\bm{b}_1$ have interpretability.  If a unit is selected into both the probability and convenience samples, it is assigned to the probability sample.  Note that the selection mechanism in (\ref{convmod}) is treated as being unknown when our methods are applied; therefore, a sample drawn in accordance with (\ref{convmod}) is a convenience sample.

We draw convenience samples under five different settings for the purpose of exploring how the inclusion mechanism influences final inferences.  The set of variables ${\bf v}_i$ and the values of the coefficients used in (\ref{convmod}) are listed in Table \ref{simconv} for each of the settings.

\begin{table}[!ht]
\centering
\footnotesize
 \caption{Coefficient values for the model used to draw the convenience sample---$\tau$ is a tuning parameter that varies the degree to which the samples are differentiated.  An asterisk indicates the variable is {\em not} used as an auxiliary variable in the calculation of weights.}
\label{simconv}
\renewcommand{\tabcolsep}{.12cm}
\hspace*{-0pt}\makebox[\linewidth][c]{%
\begin{tabular}{lrrrrr}
\hline
& \multicolumn{5}{c}{Value of coefficient in (\ref{convmod})} \\ \cline{2-6}
Variable					&	Setting 1	&	Setting 2	&	Setting 3	& 	Setting 4 	& Setting 5	\\
\hline
Intercept					&	-$\log(2)$	&	-$\log(2)$	&	-$\log(2)$	&	-$\log(2)$ 	& -$\log(2)$	\\
Caregiver depression			&	$0^*$	&	$0^*$	&	$0^*$	&	$\tau^*$ 	& $\tau^*$	\\
Caregiver anxiety				&	$0^*$	&	$0$		&	$\tau^*$	&	$0^*$ 	& $0$		\\
Caregiver gender				&	$4/3$		&	$4/3$		&	$4/3$		&	$4/3$ 		& $4/3$	\\
Caregiver age				&	$0$		&	$0$		&	$0$		&	$0$ 		& $0$		\\
Caregiver lives with care recipient	&	$1/3$		&	$1/3$		&	$1/3$		&	$1/3$ 		& $1/3$	\\
Care recipient is single			&	$1/3$		&	$1/3$		&	$1/3$		&	$1/3$ 		& $1/3$	\\
Vet.~deployed to a war zone		&	$1/3$		&	$1/3$		&	$1/3$		&	$1/3$		& $1/3$ 	\\
Vet.~has service-related disability	&	$1$		&	$1$		&	$1$		&	$1$ 		& $1$		\\
Vet.~has disability rating of 70+\%	&	$1$		&	$1$		&	$1$		&	$1$ 		& $1$		\\
Vet.~has service-related TBI		&	$1$		&	$1$		&	$1$		&	$1$ 		& $1$		\\
\hline
 \end{tabular}%
}
\end{table}

Across the settings, caregiver depression represents the outcome variable of interest and caregiver anxiety denotes a (sometimes) latent outcome variable that is strongly correlated with caregiver depression ($\rho \approx 0.65$).  The remaining variables are used as auxiliary variables in all settings.  Within Setting 1, neither depression nor anxiety influence selection into the convenience sample (although the remaining variables have a strong influence on the probability of inclusion).  Setting 2 is the same as the first, except anxiety (which does not influence the selection mechanism) is used as an auxiliary variable (this setting is considered to help assess the need for parsimony in the selection of auxiliary variables).  Setting 3 introduces circumstances where a latent outcome (anxiety) that is not the outcome of interest influences selection (note that anxiety is not used as an auxiliary variable for blending in this setting).  Setting 4 presents circumstances where the outcome of interest directly influences inclusion probabilities.  Setting 5 is identical to the fourth setting with the exception that anxiety is added to the list of auxiliary variables used for blending.  This final setting is designed to investigate whether or not one can lessen the effect of bias by including covariates that are correlated with the outcome (but do not necessarily have a direct influence on the sampling mechanisms) as auxiliary variables within the weighting schemes.  Within Settings 3, 4, and 5, the degree to which blending is insufficient (e.g., the samples are directly differentiated by depression levels) is controlled by the tuning parameter $\tau$.  Although we use $\tau = 1/2$ here, simulations are provided in the supplementary materials that consider various values of this coefficient.

A visualization of the discrepancies between the pseudo-population, the probability sample, and the convenience sample is provided in Table A.1 of the supplemental materials to this article.

Each setting of this simulation study involves $K=10,000$ independent iterations.  In each iteration, probability and convenience samples are drawn from the pseudo-population of post-9/11 military caregivers.  Each of these samples are drawn independently of one another within each iteration (although a unit that is selected for both the convenience and probability samples is assigned to the probability sample).
The four weighting methods described in Section \ref{weighting} are applied within each of the $K$ iterations of the simulation.  To review, the methods include simultaneous propensity scores (SPS), disjoint propensity scores (DPS), simultaneous calibration (SC) and disjoint calibration (DC).   The set of auxiliary variables (${\bf x}_i$) used in calculation of weights for each setting is listed in Table \ref{simconv}.

Estimated response probabilities, $\hat{r}_i$, for the probability sample are calculated using a logistic model, and the estimated probability of inclusion into the convenience sample, $d_i$, is then given by $\hat{d}_i = d_i^*\hat{r}_i$.  When the propensity score-based methods are applied, the propensity scores are calculated using the logistic model in (\ref{gammamod}).  However, $\{\gamma_i\}$ does not necessarily obey the stated logistic function under the schemes used to draw the probability and convenience samples here.  When calibration is applied, totals are estimated using the probability sample (weighted with $\hat{d}_i^{-1}$).  However, each data unit is given the same initial weight value (which is set to the number of units in the population divided by the number of unit sampled) in the calculation of calibrated weights.

To explain the bookkeeping of bias and error in parameter estimators, let $\mu$ denote the population parameter of interest, which in this case is the mean depression level of post-9/11 caregivers. Let $\hat\mu$ denote a value of $\mu$ calculated using the complete pseudo-populations ($\hat\mu$ represents a benchmark value of $\mu$).  Further, let $\hat\mu^{[k]}_j$ denote the estimated value of $\mu$ calculated using the $k^{\rm th}$ replication (for $k \in (1,\ldots,K$) and the $j^{\rm th}$ weighting scheme (for $j \in (1,\ldots,4)$).
The relative error (reported as a percent change from the benchmark) inherent in $\hat\mu^{[k]}_j$ is calculated via
$
\mbox{e}^{[k]}_j = 100(\hat\mu^{[k]}_j - \hat\mu)/\hat\mu
$
whereas the squared relative error is given by
$
\mbox{SRE}^{[k]}_j = (\mbox{e}^{[k]}_j)^2.
$ %\]
To aggregate the findings across all iterations, we approximate the bias and root-mean squared error via
\[
\mbox{bias}_j = \frac{1}{K}\sum^{K}_{k=1} \mbox{e}^{[k]}_j
~~~~\mbox{and}~~~~
\mbox{rMSE}_j = \sqrt{\frac{1}{K}\sum^{K}_{k=1} \mbox{SRE}^{[k]}_j}.%,
\]

Let $p^{[k]}_j$ denote the $p$-value of a test of the hypotheses in (\ref{h0}) (or related hypotheses if a more exhaustive regression model is used) for the $k^{\rm th}$ replication with the $j^{\rm th}$ weighting scheme.  We report
\[
\hat{p}_j = \frac{1}{K}\sum^{K}_{k=1} 1_{\{p^{[k]}_j \leq \alpha \}}
\]
which denotes the rate at which the null hypothesis of adequate blending is rejected.  Recall that $1_{\{A\}}$ denotes the indicator of event $A$.  We use a significance level of $\alpha = 0.05$.  The rejection rate $\hat{p}_j$ approximates type I error in the event that ${\cal H}_0$ is true (e.g., Settings 1 and 2) and the power of the test when ${\cal H}_1$ is true (e.g., Settings 3--5).
We also calculate the design effect that is seen in $\hat{\mu}$ and we calculate the standard error of this quantity; to avoid redundancy, we only report the design effect (the comparative results for the standard errors are similar).
Lastly, we calculate the {\em post hoc} blending estimator from (\ref{posthoc}) when using the two sets of disjoint weights.  To approximate the variances and covariances in (\ref{barkap}) for the {\em post hoc} estimator, we use a jackknife.  Only the bias and rMSE are reported for the {\em post hoc} estimator of $\mu$.
Standard errors and the design effect corresponding to $\hat\mu$ are calculated using Taylor series linearization.
To ensure computational feasibility of these simulations, we do not use resampling approaches.

Findings are reported in Table \ref{simresults} for $\tau=1/2$.  Within each of the settings, weighted blending (regardless of method) is preferable to unweighted blending.  As expected, if the discrepancies between the probability and convenience samples have been appropriately modeled (i.e., Setting 1 and 2), weighted blending produces more accurate (lower bias) and more precise (lower rMSE) estimators than those which are found using only the probability sample; otherwise (i.e., Settings 3--5), it may be preferable to use only the probability sample.  It appears that calibration and propensity scores produce similar results.  However, we note that the propensity score-based methods tend to yield lower design effects, whereas the calibration methods often observe lower rMSE.  Further, we see that simultaneous weighting yields lower design effects and consequentially lower rMSE than disjoint weighting methods.  We also see that when blending is based on a sufficient model (i.e., Setting 1), the rejection rate of the hypotheses in (\ref{h0}) is close to its nominal level so long as disjoint blending is used; also, our expectation is validated that simultaneous weights are inappropriate for testing the sufficiency of blending.

\begin{table}[!ht]
\centering
\small
\renewcommand{\tabcolsep}{.13cm}
 \caption{Results under the simulated settings of Table \ref{simconv} when $\tau = 1/2$.  Methods used include the probability sample only (KP), unweighted blended samples (unw), simultaneous propensity scores (SPS), disjoint propensity scores (DPS), simultaneous calibration (SC), and disjoint calibration (DC).  Results are also reported for the {\em post hoc} blending estimator when disjoint propensity score weights ($\bar\kappa$PS) and disjoint calibration weights ($\bar\kappa$C) are used.  Results regard the estimator of ${\mu}$, the mean depression level in the pseudo-population of post-9/11 caregivers.  Each setting uses $K=10,000$ iterations.}   \label{simresults}
\vspace*{-0.0in}\makebox[\linewidth][c]{%
\begin{tabular}{clcccccccc}
\hline
 & & KP & unw & SPS & SC & DPS & DC & $\bar\kappa$PS & $\bar\kappa$C\\
\hline
\multirow{4}{*}{Setting 1} & DEFF & 1.00 & 1.00 & 1.71 & 1.88 & 1.83 & 2.32 & --- & --- \\
& Bias & 5.70 & 13.30 & 0.56 & -0.02 & 1.98 & 0.28 & 1.72 & -0.16 \\
& rMSE & 11.53 & 13.55 & 5.71 & 5.98 & 6.27 & 6.54 & 6.31 & 6.67 \\
& Rej.~rate & --- & 0.14 & 0.17 & 0.19 & 0.05 & 0.04 & --- & --- \\ \hline
\multirow{4}{*}{Setting 2} & DEFF & 1.00 & 1.00 & 1.72 & 1.89 & 1.84 & 2.33 & --- & --- \\
& Bias & 5.80 & 13.32 & 0.34 & -0.29 & 1.72 & -0.23 & 1.48 & -0.57 \\
& rMSE & 11.41 & 13.56 & 7.18 & 7.56 & 7.55 & 7.82 & 7.55 & 7.93 \\
& Rej.~rate & --- & 0.13 & 0.15 & 0.18 & 0.02 & 0.01 & --- & --- \\ \hline
\multirow{4}{*}{Setting 3} & DEFF & 1.00 & 1.00 & 1.75 & 1.91 & 1.86 & 2.33 & --- & --- \\
& Bias & 5.72 & 19.89 & 8.49 & 7.63 & 10.64 & 8.53 & 10.25 & 8.09 \\
& rMSE & 11.47 & 20.05 & 10.19 & 9.68 & 12.19 & 10.83 & 12.06 & 10.74 \\
& Rej.~rate & --- & 0.33 & 0.46 & 0.46 & 0.23 & 0.18 & --- & --- \\ \hline
\multirow{4}{*}{Setting 4} & DEFF & 1.00 & 1.00 & 1.78 & 1.94 & 1.91 & 2.39 & --- & --- \\
& Bias & 5.80 & 24.06 & 13.86 & 12.91 & 16.50 & 14.32 & 15.57 & 13.35 \\
& rMSE & 11.58 & 24.19 & 15.01 & 14.29 & 17.61 & 15.91 & 17.06 & 15.47 \\
& Rej.~rate & --- & 0.50 & 0.66 & 0.66 & 0.47 & 0.41 & --- & --- \\ \hline
\multirow{4}{*}{Setting 5} & DEFF & 1.00 & 1.00 & 1.76 & 1.94 & 1.90 & 2.43 & --- & --- \\
& Bias & 5.67 & 24.01 & 8.88 & 7.78 & 11.67 & 8.36 & 11.17 & 7.82 \\
& rMSE & 11.48 & 24.15 & 11.60 & 11.04 & 14.05 & 11.85 & 13.85 & 11.74 \\
& Rej.~rate & --- & 0.50 & 0.64 & 0.66 & 0.23 & 0.12 & --- & --- \\ \hline
\hline
\end{tabular}%
}
\end{table}

From Setting 2, we learn that parsimony in the choice of auxiliary variables is beneficial since the second setting observes higher rMSE than the first (likewise, the test for the adequacy of blending is conservative in the Setting 2).  However, the remaining settings illustrate that it is necessary to have a robust set of auxiliary variables.
That is, even though anxiety is not the outcome of interest, we see in Setting 3 that allowing the probability of selection into the convenience sample to depend upon anxiety induces bias into estimators found by the weighting schemes.  This bias would be smaller if depression and anxiety were not highly correlated.  Similarly, Setting 5 (when compared to Setting 3) illustrates that if the probability of selection into the convenience sample depends on the outcome of interest (depression), bias can be reduced by using additional variables that are correlated with the outcome as auxiliary variables in the calculation of weights.

From Table \ref{simresults}, we also see that the {\em post hoc} blending estimator performs comparably to the corresponding estimators founding using the corresponding disjoint blending weights in all settings.  Therefore, we conclude that there is little to no loss of efficiency that stems from the use of the approximated design effect in our disjoint blending technique in comparison to more rigorous methods of variance minimization.

The supplemental materials present further simulations in this setup that study the effect of the parameter $\tau$ on the performance of the methods.

\subsection{Simulations with synthetic data} \label{synthetic}

Here, we compare methods for variance estimation with blended data using synthetic data.
In these simulations, we vary the degree to which auxiliary variables used for blending are correlated with the outcome of interest.  For a population of size $N=10,000$, we generate a 2-dimensional vector of auxiliary variables via ${\bf X} = (X_1,X_2,X_3) \sim N({\bf 0},{\bf I}_3)$, where ${\bf I}_k$ is the identity matrix of dimension $k$.  The outcome is simulated using $Y = \beta (X_1 + X_2) + \epsilon$, where $\epsilon \sim N(0, \sigma^2_e)$.  We choose $\beta$ and $\sigma_e^2$ so as to ensure the coefficient of determination, $R^2 = 2\beta^2/(2\beta^2 + \sigma_e^2)$, takes on a desired value while maintaining $\mbox{Var}(Y)=1$.  That is, for a given value of $R^2$, we set $\beta = \sqrt{R^2/2}$ and $\sigma_e^2 = 1-R^2$.  Next, we select $n_1 = 200$ case for the probability sample at random from the simulated population.  Those cases are considered respondents (and consequentially elements of the probability sample) with probability $r_i$ that satisfies $r_i = 1/\{1+\exp(-0.15X_3)\}$. An element of the simulated population is assigned to the convenience sample with a probability determined by $p = 1/\{1+\exp[-4.2 - 0.5(X_1 + X_2)]\}$.  As before, data elements selected into both samples are assigned to the probability sample.  Blending weights are then calculated using simultaneous propensity scores.  Logistic modeling is used to estimate response probabilities and propensity scores as functions of $(X_1,X_2)$ and $X_3$, respectively.

Letting $\eta = \mbox{E}[Y]$, we calculate $\hat\eta$, the estimate of $\eta$, when found using only the probability sample and when found using the (weighted) blended sample.  We also calculate the standard error of $\hat\eta$ when only the probability sample is used,
when the blended weights are used with Taylor series linearization for variance estimation,
and when a delete-a-group jackknife (as outlined in Section \ref{var}) with $G=40$ is used.
Note that weights are recalculated for each replicate group in the jackknife.  Using the estimated standard error (and the corresponding value of $\hat\eta$), we can calculate the upper and lower bounds of a $95\%$ confidence interval.  For each confidence interval, we track the percent of iterations in which the true mean, $\eta = 0$, falls in the interval---this is referred to as coverage.

\begin{figure}[!htb]
\centering
\hspace*{-12pt}\makebox[\linewidth][c]{%
\begin{tabular}{cc}
\includegraphics[width=2.4in, bb = 4 19 330 302]{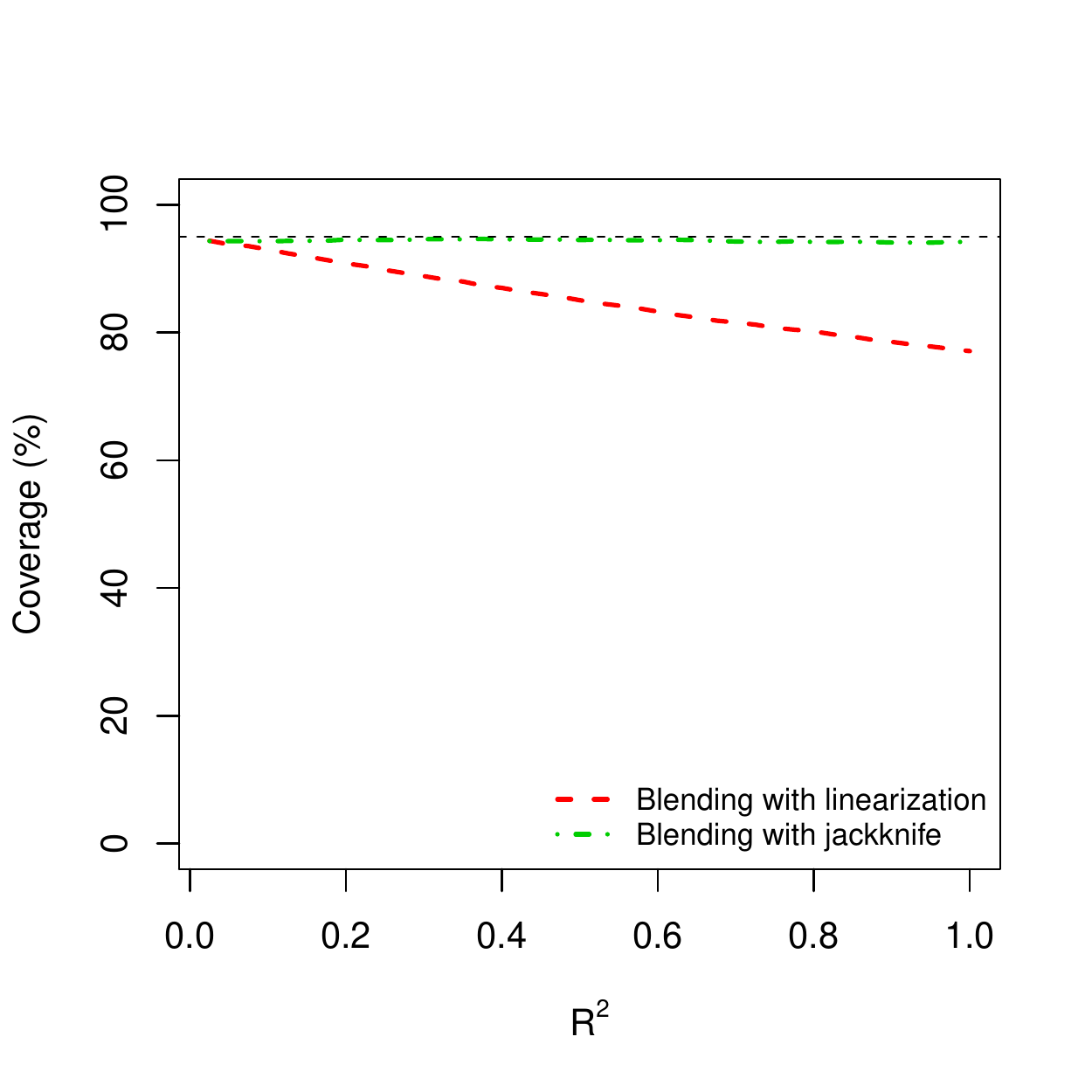}
&
\includegraphics[width=2.4in, bb = 4 19 330 302]{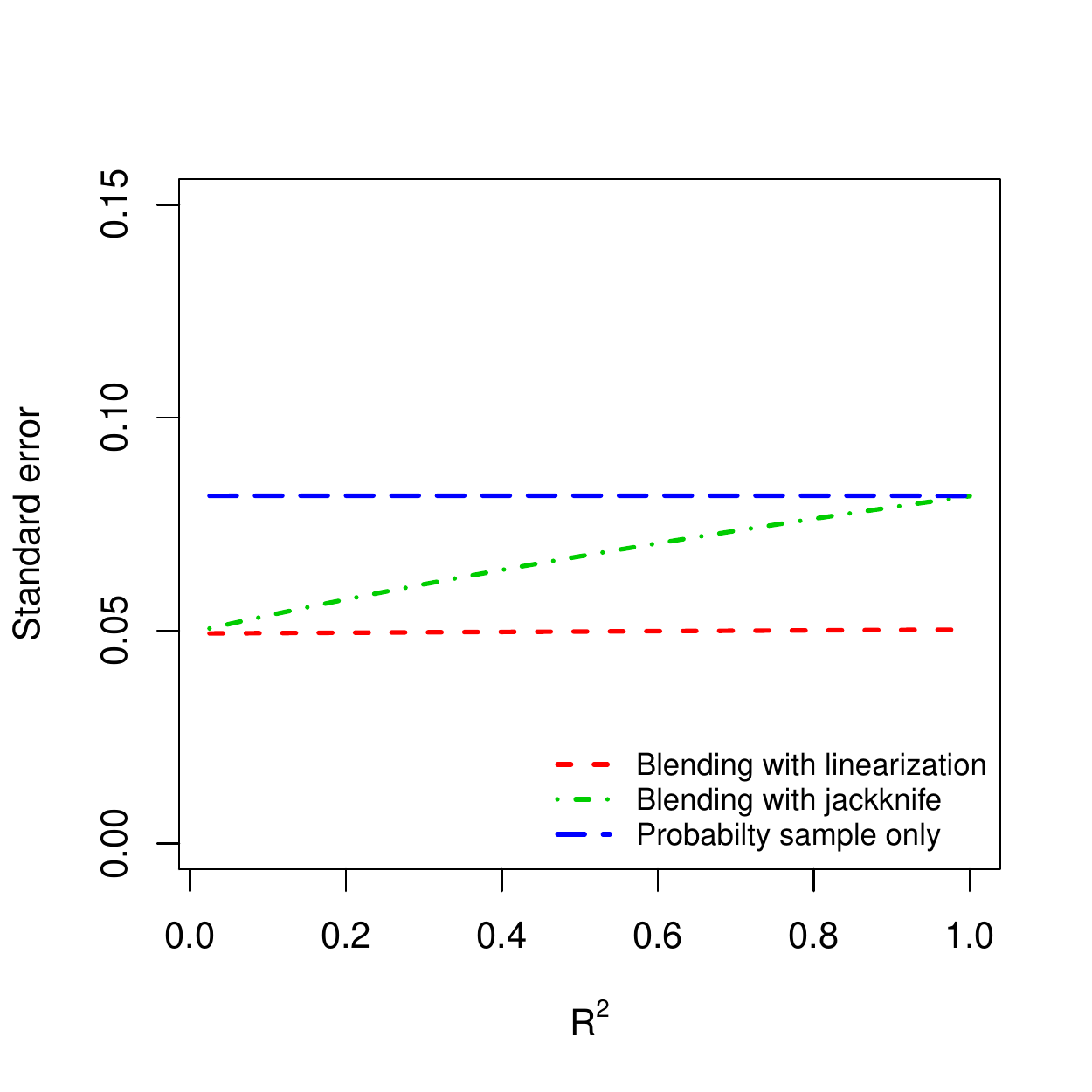}
\end{tabular}%
}
\caption{Coverage (left) and standard errors (right) in the estimate of $\eta = \mbox{E}[Y]$ as a function of $R^2$ for various methods of variance estimation.  Blending is performed using simultaneous propensity scores.  For each value of $R^2$, 10,000 iterations are used.}
\label{fig2}
\end{figure}

Results are shown in Figure \ref{fig2}.  We see that coverage remains at the nominal level of 95\% when variance estimation is performed with a jackknife; however, coverage decreases with increasing $R^2$ when linearization is used.  Therefore, one should not estimate variance with linearization when the auxiliary variables are moderately to strongly related with the outcome.  Further, in this study, standard error remains constant across all values of $R^2$ when only the probability sample is used and when blending is performed with variance estimated using linearization.  However, standard error increases with $R^2$ when the variance of the blending estimator is calculated with a jackknife.  In fact, when $R^2 = 0$, the jackknife yields the same efficiency as linearization; however, when $R^2=1$, the jackknife yields a standard error equivalent to that provided by only the probability sample.
Therefore, if one were to use the outcome as an auxiliary variable for blending, one would have little to no gain in precision from blending samples over what would be achieved with only the probability sample.  These observations imply that the design effect and rejection rates reported for Settings 2 and 5 in Table \ref{simresults} are likely understated since linearization was used (in Settings 2 and 5, $R^2 = 0.42$; otherwise, $R^2 = 0.14$).

\section{Discussion} \label{conclude}

We introduced four methods for calculating weights that blend probability and convenience samples (using combinations of disjoint vs.~simultaneous blending and propensity scores vs.~calibration).  Simultaneous methods appear to yield lower design effects (and variances).  In addition, simultaneous blending does not require coverage of the convenience sample (i.e., Assumption \ref{assump4}). However, disjoint blending is needed to assess the adequacy of the set of auxiliary variables.  Calibration and propensity score procedures perform comparably, with propensity scores yielding slightly smaller design effects in our empirical illustrations.  Note that calibration requires a feasible solution to the calibration equations (which was not satisfied for disjoint calibration in our application).  For variance estimation, a jackknife was shown to outperform linearization in simulations.

We revisit the assumptions mandated by our approaches and discuss the degree to which they were satisfied in our empirical examples.  With respect to Assumptions \ref{assump1} and \ref{assump1a} in the caregiver data, the probability sample appears to cover the population; however, as noted in Section \ref{calcprop}, the proprietary nature of the KP prevented us from rigorously adjusting for non-response when estimating $d_i$ for the convenience sample.  Our solution to this issue was to assume that KP non-respondents are not substantially different from respondents, which appears reasonable observationally.  Further, $d_i$ is not needed for all $i \in S$ for calibration methods, and propensity score methods provided results that were similar to calibration in the military caregivers application.  Assumption \ref{assump2} appears upheld in our data application since the test for the adequacy of blending did not reject for any outcomes.  Assumption \ref{assump3} (adequacy of the propensity score model) may not have been upheld.
In fact, in our simulation studies, it was not correct (since a logistic model was used to draw the convenience sample, the propensity score model will have a different form).  The simulations showed that despite this, the propensity score-based techniques perform well.
Lastly, although the WWP was strongly differentiated from the KP, we saw no evidence to suggest that Assumption \ref{assump4} (positivity of the convenience sample) was not satisfied.

Although we made efforts within our simulation study to provide guidance regarding how one should select a choice of auxiliary variables that are to be used in the methods outlined here, we feel that more work in this vein is needed. For instance, \cite{brookhart06} provide a thorough assessment of model selection within traditional propensity score procedures.  They suggest that variables that are not related to treatment status but related to the outcome should always be included.  However, this guidance does not translate to our work:~on account of the findings of Section \ref{synthetic}, building propensity scores for blending while unnecessarily using auxiliary variables that are strongly related to the outcome will lead to a reduction in the gain in precision provided by the convenience sample.  Therefore, we recommend being parsimonious when selecting auxiliary variables.

\setlength{\bibsep}{0.0pt}
\singlespacing
\bibliographystyle{Chicago}

\bibliography{big_biblio}

\end{document}